\documentclass[slac_one]{revtex4}
\usepackage{graphicx}
\usepackage{fancyhdr}
\newcommand{\noi}{\noindent}
\newcommand{\be}{\begin{equation}}
\newcommand{\ee}{\end{equation}}
\newcommand{\beq}{\begin{eqnarray}}
\newcommand{\eeq}{\end{eqnarray}}
\newcommand{\etal}{{\it et al}.,}

\newcommand{\sinsthzw}{\sin^2\theta^0_W}
\newcommand{\sinsthw}{\sin^2\theta_W}
\newcommand{\sinshad}{\sin^2\theta_W(m_Z)^{\rm hadronic}_{\overline{MS}}}
\newcommand{\sinslep}{\sin^2\theta_W(m_Z)^{\rm leptonic}_{\overline{MS}}}
\newcommand{\mzms}{(m_Z)_{\overline{MS}}}
\newcommand{\stmod}{SU(3)$_C\times{}$SU(2)$_L\times{}$U(1)$_Y$}
\newcommand{\stmodbf}{SU(3)$_\mathbf{C}\mathbf{\times}{}$SU(2)$_\mathbf{L}\mathbf{\times}{}$U(1)$_\mathbf{
Y}$}

\newcommand{\NC}{{\em Nuovo Cimento}}

\newcommand{\NP}{{\em Nucl. Phys.}}
\newcommand{\PL}{{\em Phys.\ Lett.}}
\newcommand{\PRL}{{\em Phys.\ Rev.\ Lett.}}
\newcommand{\PR}{{\em Phys.\ Rev.}}

\newcommand{\EPJ}{{\em Eur.\ Phys.\ J.}}
\newcommand{\ARNPS}{{\em Annual Review of Nucl.\ and Part.\ Sci.}}
\newcommand{\RMP}{{\em Rev.\ Mod.\ Phys.}}

\newcommand{\lsim}{\mathrel{\lower4pt\hbox{$\sim$}}
\hskip-12.5pt\raise1.6pt\hbox{$<$}\;}

\newcommand{\gsim}{\mathrel{\lower4pt\hbox{$\sim$}}
\hskip-11.5pt\raise1.6pt\hbox{$>$}\;}

\begin{document}

\title{Precision Electroweak Measurements and the Higgs Mass\footnote{2 Lectures given at the XXIII SLAC Summer Institute, SSI 2004 ``Natures Greatest Puzzles'', Aug.~2--13, 2004.}}

%

\author{William J. Marciano}
\affiliation{Brookhaven National Laboratory \\ Upton, New York\ \ 11973}

\begin{abstract}
The utility of precision electroweak measurements for predicting the Standard Model Higgs mass via quantum loop effects is discussed. Current constraints from $m_W$ and $\sinsthw\mzms$ imply a relatively light Higgs $\lsim 154$ GeV which is consistent with Supersymmetry expectations. The existence of Supersymmetry is further suggested by a discrepancy between experiment and theory for the muon anomalous magnetic moment. Constraints from precision studies on other types of ``New Physics'' are also briefly described.
\end{abstract}

\maketitle

\thispagestyle{fancy}


\noi {\bf Outline}

\noi 1. Higgs Mass and Precision Measurements --- A Preamble

\noi 2. The Standard \stmod\ Model

\noi 3. Natural Relations: $\sinsthzw =  e^2_0/g^2_{2_0} = 1-(m^0_W/m^0_Z)^2$

\noi 4. Renormalized Parameters

4.1. $\alpha$

4.2. $G_\mu$

4.3. $m_Z$ and $m_W$

4.4. $\sinsthw\mzms$

\noi 5. Electroweak Radiative Corrections

5.1. $m_t$ and $m_H$ Dependence

\noi 6. Higgs Mass Prediction

6.1. $S$ Parameter Constraints

\noi 7. The Muon Anomalous Magnetic Moment

7.1. The $a_\mu$-$m_H$ Connection

\noi 8. Other Precision Studies

8.1. SLAC E158: Polarized $e^-e^-$ Asymmetry

\noi 9. Outlook and Conclusion

\section{Higgs Mass and Precision Measurements --- A Preamble}

It has been known for some time that Standard Model quantum loops exhibit a small but important dependence on the Higgs mass, $m_H$ \cite{npb84132,prd20274}. As a result, the value of $m_H$ can, in principle, be predicted by comparing a variety of precision electroweak measurements with one another. Toward that end, recent global fits to {\it all\/} precision electroweak data (see J. Erler and P. Langacker \cite{jerlerref4}) give

\beq
m_H & = & 113^{+56}_{-40} {\rm ~GeV} \label{eqone} \\
m_H & < & 241 {\rm ~GeV}\qquad (95\% {\rm~CL}) \label{eqtwo}
\eeq

\noi Those constraints are very consistent with bounds \cite{plb5921} from direct searches for the Higgs boson at LEPII via $e^+e^-\to ZH$

\be
m_H > 114.4 {\rm ~GeV} \label{eqthree}
\ee

\noi Together, they suggest the range 114 GeV${}< m_H<{}$241 GeV, and imply very good  consistency between the minimal Standard Model theory and experiment.

Global fits \cite{prd361385} are very useful, when many different measurements of similar precision are included. However, sometimes it is instructive to be subjective, particularly with regard to systematic errors including theory uncertainties. Global fits, if blindly accepted, may be washing out interesting aspects of the data. It is with that point of view that I approach these lectures. We have a subset of very clean precise measurements that can on their own overconstrain the Standard Model and be used to predict the Higgs mass and/or search for ``New Physics''. Concentrating on those measurements instead of the global fit allows for a more transparent discussion of the $m_H$ sensitivity. It also suggests, as we shall see, a somewhat lighter Higgs and possibly the advent of supersymmetry (if you stretch your imagination).

Having advised the reader of my subjectivity, let me briefly discuss the input I use and the reasons for my prejudice. First, there are the very precisely measured electroweak parameters $\alpha$, $G_\mu$ (Fermi Constant) and $m_Z$. Their values (within errors) are unquestioned. I then compared those quantities with $m_W$ and $\sinsthw\mzms$; with the latter  extracted only from leptonic $Z$ decay asymmetries. Then, I use the recently updated value $m_t=178$ GeV as input in loops. The resulting comparison overdetermines $m_H$ within the Standard Model framework. It is encouraging, however, that $m_W$ and $\sinslep$ together point to $m_H\lsim 154$ GeV and provide a nice consistency check on one another.

My preference for $\sinslep$ over $\sinshad$ (obtained from $Z\to{}$hadrons) or an average of the two may be viewed as controversial; so, let me elaborate somewhat on that choice.

\begin{enumerate}

\item LEPII and Tevatron determinations of $m_W$ are consistent and can be averaged with some confidence. The value of $\sinslep$ is consistent with $m_W$ (within the Standard Model framework) while $\sinshad$ is not.

\item Extraction of $\sinslep$ from $A_{LR}$ at the SLC is theoretically pristine and consistent with the value obtained at LEP using $A_{FB}(Z\to\ell^+\ell^-)$. That situation is to be contrasted with $\sinshad$ extracted (with high statsistics) from $A_{FB}(Z\to b\bar b)$ at LEP which disagrees with $A^{LR}_{FB} (Z\to b\bar b)$ at th SLC\null. Although $A_{FB}(Z\to b \bar b)$ has been thoroughly scrutinized experimentally, one has the feeling that some theoretical or systematic effect that could shift $\sinshad$ may have been overlooked. That view is partly based on the history of $Z\to b\bar b$ studies where an anomaly in $\Gamma(Z\to b\bar b)$ came and went.

\end{enumerate}

Fortunately, we can expect $m_W$ to improve somewhat in the near term at the Tevatron and in the longer term at the LHC\null. It will be difficult to improve $\sinslep$ without a $Z$ factory or very intense high energy linear collider (see also the M\o ller scattering discussion in Section~8). Perhaps $\sinshad$ will be reexamined. Indeed, there seems to be some shift from recent $A_{FB}(Z\to b\bar b)$ studies by the DELPHI collaboration \cite{epjc34127}. Otherwise, it is hard to see how the discrepancy in $\sinsthw\mzms$ will be resolved.

Having expressed my partisanship, let me describe the goals and contents of these two talks. I will try to explain how precision electroweak measurements are used to constrain the Higgs mass, $m_H$, and some forms of new physics. That discussion will entail a brief description of the \stmod\  Standard Model in Section~2 and some natural relations among bare (unrenormalized) couplings and masses in Section~3. Those natural relations stem from a custodial isospin symmetry that overconstrains the Standard Model in a way that provides sensitivity to quantum loops and the Higgs mass. Then in section~4, I review the definition and  status of some precision electroweak parameters. Their interconnection and dependence on the top quark mass, $m_t$, and $m_H$ is described in Section~5. That is followed by Section~6 which provides a Higgs mass prediction based on quantum loops along with constraints on the Peskin-Takeuchi \cite{prl65964} new physics parameter, $S$. I switch gears in Section~7 and review the status \cite{arnps54,marcpas04} of the muon anomalous magnetic moment, $a_\mu$, emphasizing the current discrepency between theory and experiment and its possible interpretation as a sign of supersymmetry. Then I briefly discuss several other precision low energy experiments with primary focus on the recently completed E158 (polarized M\o ller scattering \cite{kolossi04}) at SLAC and the running weak mixing angle. Finally, in Section~9 an outlook on the future along with some concluding remarks are given. 

\section{The Standard \stmodbf\ Model}

The Standard \stmod\ Model of strong and electroweak interactions has been enormously successful. Based on the principle of local gauge invariance, it follows the modern approach to elementary particle physics in which ``Symmetry Dictates Dynamics''. Amazingly, the SU(3)$_C$ symmetry of Quantum Chromodynamics (QCD) describes all of strong interaction physics via simple quark-gluon interactions. On its own, QCD has no free parameters \cite{pr36137}. However, if a unit of mass is introduced via electroweak physics (e.g.\ $m_e$), then the QCD coupling becomes the single pure QCD parameter and its value is found to be (at scale $m_Z=91.1875$ GeV) \cite{plb5921}

\be
\alpha_s(m_Z) \equiv \frac{g^2_s(m_Z)}{4\pi} = 0.119(2)\qquad {\overline{MS}} {\rm ~definition} \label{eqfour}
\ee

The SU(2)$_L\times{}$U(1)$_Y$ sector is much more arbitrary \cite{marcssi93}. Depending on ones counting, it has at least 24 independent parameters. They include: 2 bare gauge couplings $g_{2_0}$ and $g_{1_0}$ (usually traded in for $\tan\theta^0_W = \sqrt{3/5} g_{1_0}/g_{2_0}$ and $e_0=g_{2_0}\sin\theta^0_W$), 2 Higgs potential parameters $\lambda_0$ (the self coupling) and $v_0$ (vacuum expectation value) and 36 complex Yukawa couplings connecting the Higgs doublet and 3 generations of quarks and leptons. Of the 72 Yukawa coupling (real) parameters, only 20 are observable as quark and lepton masses and mixing (phase) angles. Other possibilities include $\bar\theta$ (a QCD enhanced CP violating parameter), 2 relative phases in the case of Majorana neutrinos, and right-handed neutrino mass scales if a see-saw mechanism for neutrino masses is adhered to.

A goal of experimental physics is to measure the electroweak parameters as precisely as possible while trying to uncover new physics or deeper insights. Theoretical studies aim to refine or better understand Standard Model predictions while also exploring ideas for physics beyond the Standard Model. The latter include additional symmetries such as grand unification, supersymmetry, extra dimensions etc. Symmetries can, in principle, reduce the number of arbitrary electroweak parameters by promoting natural relations among otherwise independent quantities (see Section~3). Ultimately, one aims for a parameter free description of Nature, a noble but difficult goal.

Tests of the Standard Model have been extremely successful. They entail 25 years of discovery and precision tests. Collectively, they have uncovered all Standard Model gauge bosons and 3 generations of fermions. In addition, measurements at the $\pm0.1\%$ or better level have tested quantum loop effects. What remains elusive is the so-called Higgs scalar particle, $H$, a remnant of the fundamental Higgs mechanism responsible for electroweak mass generation.

Let me say a few words about the minimal Higgs mechanism. It is based on the introduction of a complex scalar doublet, $\phi$, to the fundamental Lagrangian

\be 
\phi(x)= \frac{1}{\sqrt{2}} \left( {w_1(x) + i w_2(x)} \atop {H+v_0+iz(x)} \right) \label{eqfive}
\ee

\noi via the appendage of the potential $V(\phi)$

\be 
V(\phi) = \lambda_0(\phi^+\phi -v^2_0/2)^2 \label{eqsix}
\ee

\noi which breaks the SU(2)$_L\times{}$U(1)$_Y$ electroweak symmetry down to U(1)$_{em}$ of electromagnetism.

Parts of the scalar doublet $w^\pm = \frac{1}{\sqrt{2}} (w_1 \pm iw_2)$ and $z$, are would-be massless goldstone bosons. They become longitudinal components of the $W^\pm$ and $Z$ gauge bosons, endowing them with masses. So, in a sense, 3/4 of the scalar doublet has been discovered (via $W^\pm$ and $Z$ discovery). Of course, the remaining 1/4 is the remnant physical Higgs scalar, $H$. It represents an important missing link of the Higgs mechanism and must be discovered before one can be confident that the Standard Model is correct (even at an effective low energy level).

Direct searches for the Higgs via $e^+e^-\to ZH$ at LEPII provide the experimental bound in eq.~(\ref{eqthree}) ($m_H\gsim114$ GeV), while global fits to precision measurements suggest a ``best bet'' value in the vicinity of that bound (see eq.~(\ref{eqone})). The Standard Model predicts (I use lowest order or bare parameters)

\be
m^0_H = \sqrt{2\lambda_0} v_0 \simeq \sqrt{\lambda_0}\times 350 {\rm ~GeV} \label{eqseven}
\ee

\noi where

\be
v_0 = 2m^0_W/g_{2_0} \simeq 250 {\rm ~GeV} \label{eqeight}
\ee

\noi So, a strong coupling Higgs scalar sector, $\lambda\gsim 1$ corresponds to a relatively heavy $m_H\gsim 350$ GeV, while weak coupling $\lambda\lsim e$ implies a relatively light $m_H\lsim 190$ GeV\null. Given the lack of successs of dynamical symmetry breaking schemes where effectively $\lambda>1$ and the popularity of supersymmetry where $\lambda$ is weak, a light Higgs is theoretically and phenomenologically preferred.

Let me make a few observations about Standard Model masses. First, there is the Higgs vacuum expectation value $v_0 \simeq 250$ GeV\null. It is the scale of electroweak symmetry breaking. All electroweak masses are proportional to $v_0$, a quantity that exhibits quadratic divergencies (mass ratios are free of quadratic divergencies). The quadratic divergence is not necessarily a fundamental flaw in the Standard Model (it is renormalized away). However, when the Standard Model is embedded in a high mass scale theory (e.g.\ grand unified, superstring etc.) a mass hierarchy results which requires fine-tuning if it is to be maintained order by order in perturbation theory. That failing suggests new physics which ameliorates the need for fine tuning. It is a major stimulus for supersymmetry.

The relative strengths of the electroweak gauge couplings are parametrized by the weak mixing angle $\theta_W$

\be
\tan\theta^0_W \equiv \sqrt{3/5} g_{1_0}/g_{2_0} \label{eqnine}
\ee

\noi (It is sometimes noted as a historical remark that the $W$ in $\theta_W$ corresponds to either the first letter in Weinberg \cite{prl191264} or last letter in Glashow \cite{np22579}.) The weak mixing angle is very fundamental and appears in many contexts. It is of primary importance in SU(2)$_L-{}$U(1)$_Y$ mixing. The $W^\pm$ mass is given by 

\be
m^0_W = \frac12 g_{2_0} v_o \label{eqten}
\ee

\noi while the $W_3$-$B$ gauge boson mixing mass matrix squared is given by

\be
M^2 \sim \left(\begin{array}{cc}
g^2_{2_0} & \sqrt{3/5} g_{1_0}g_{2_0} \\
\sqrt{3/5} g_{1_0}g_{2_0} & 3/5 g^2_{1_0}\end{array}
\right) v^2_0 \label{eqeleven}
\ee

\noi (Note, I assume a normalization $g^\prime_0=\sqrt{3/5} g_{1_0}$, where $g^\prime_0$ is often employed as the U(1)$_Y$ coupling.) Diagonalization of the mass matrix leads to the neutral gauge boson eigenstates.

\beq
Z & = & W_3\cos\theta^0_W - B\sin\theta^0_W \qquad m^0_Z = m^0_W/\cos \theta^0_W \label{eqtwelve} \\
\gamma & = & B\cos\theta^0_W + W_3\sin\theta^0_W \qquad m_\gamma=0 \label{eqthirteen}
\eeq 

\noi parametrized by $\theta^0_W$. The photon coupling to charged particles requires

\be
e_0 = g_{2_0}\sin\theta^0_W \label{eqfourteen}
\ee

\noi  I have used bare quantities in all relations. The Standard Model is a renormalizable quantum field theory; so, its parameters undergo (infinite and finite) renormalization. In terms of the renormalized quantities, all phenomena are finite and calculable. However, as we shall now see, the renormalization is constrained by symmetries, a fortunate feature that gives us a handle on $m_H$ and possible new physics.

\section{Natural Relations: $\mathbf{\sinsthzw=e^2_0/g^2_{2_0} =1-(m^0_W/m^0_Z)^2}
$}

Sometimes, due to a symmetry, two parameters are related at the bare level, such that the same relationship is maintained at the renormalized level, up to finite calculable radiative corrections. When that is the case, the relationship is called natural. Let me give a few simple examples.
\medskip

\noi {\it i}) Electron-Muon-Tau Universality:  All lepton doublets have the same SU(2)$_L$ coupling, $g_{2_0}$, due to local SU(2)$_L$ gauge invariance. Therefore, all $W\ell \nu$ renormalized couplings differ from $g_{2_0}$ by the same infinite renormalization \cite{prd83612}. Hence, ratios such as $\Gamma(W\to e\nu)/\Gamma(W\to\mu\nu)$ etc.\ are finite and calculable to all orders in perturbation theory. Such relations have been well confirmed with high precision \cite{npb403} in many weak decays of the $W^\pm$, $\tau^\pm$, $\pi^\pm$ etc.
\medskip

\noi {\it ii}) CKM Unitarity: Unitarity among CKM quark mixing parameters requires $\sum_k V^0_{ik} V^{0^*}_{jk} = \delta_{ij}$ for the bare matrix elements. So, for example, the first row should satisfy

\be
|V^0_{ud}|^2 + |V^0_{us}|^2 + |V^0_{ub}|^2 = 1 \label{eqfifteen}
\ee

\noi The infinite renormalizations of those 3 quantities are naturally related such that the measured parameters satisfy

\be
|V_{ud}|^2 + |V_{us}|^2 + |V_{ub}|^2 = 1 +\delta \label{eqsixteen}
\ee

\noi where $\delta$ is finite and calculable. In fact, one can and usually does define the renormalized mixing parameters such that $\delta=0$. Then if measurements suggest a violation of unitarity, it implies new physics beyond the Standard Model. Currently no deviation is seen, but the value of $|V_{us}|$ is still somewhat controversial \cite{hepph0406324}.
\medskip

\noi {\it iii}) Fermion Masses $m^0_b=m^0_\tau$:  In the Standard Model $m^0_b$ and $m^0_\tau$ (and all other fermion masses) are independent.  Each undergoes a different infinite renormalization. However, in some grand unified theories (GUTS), the relationship $m^0_b=m^0_\tau$ can be natural. Divergencies from gauge boson and scalar loop corrections turn out to be the same for both, but there are large finite corrections from logarithmically enhanced effects. In that way $m^0_b/m^0_\tau=1$ becomes $m_b/m_\tau\simeq 2.5$ (as roughly observed) at the physical (or renormalized) mass level. In GUTS, another famous natural relationship is $g_{3_0}=g_{2_0} =g_{1_0}$, which gives insight regarding the scale of unification $m_X\simeq10^{16}$ GeV\null.

Natural relations among bare parameters is clearly a powerful constraint, particularly when the quantities involved appear to be so different (e.g.\ $m^0_b=m^0_\tau$). In the Standard Model, there is a custodial global SU(2)$_V$ isospin like symmetry that is preserved by the simple Higgs doublet symmetry breaking mechanism. It gives rise to the natural relationships \cite{nc16a423}

\be
\frac{e^2_0}{g^2_{2_0}} = 1- (m^0_W/m^0_Z)^2 = \sinsthzw \label{eqseventeen}
\ee

\noi mentioned in Section~2. Eq.~(\ref{eqseventeen}) is quite amazing. It relates gauge boson masses, couplings and the weak mixing angle. Each of the 3 quantities in eq.~(\ref{eqseventeen}) exhibit the same ultraviolet divergencies. However, they have different finite radiative corrections \cite{npb84132}. Those finite part differences are sensitive to fermion loop effects, $m_t$, $m_H$ and potential new physics effects via loop or tree level effects.

Using the bare Fermi constant

\be
G^0_\mu = \frac{g^2_{2_0}}{4\sqrt{2}m^{0^2}_W} \label{eqeighteen}
\ee

\noi one can recast eq.~(\ref{eqseventeen}) into the forms

\beq
G^0_\mu & = & \frac{\pi\alpha_0}{\sqrt{2} m^{0^2}_W (1-m^{0^2}_W/m^{0^2}_Z)} = \frac{\pi\alpha_0}{\sqrt{2} m^{0^2}_W\sinsthzw} \nonumber \\
& = & \frac{2\sqrt{2}\pi\alpha_0}{m^{0^2}_Z\sin^22\theta^0_W} \label{eqnineteen}
\eeq

\noi Those same relations hold among renormalized parameters, up to finite calculable corrections. Of course, the actual finite corrections will depend on the exact definitions of renormalized parameters employed. So, for example, one expects \cite{prd22471}

\be
G_\mu (1-\Delta r) = \frac{\pi\alpha}{\sqrt{2} m^2_W(1-m^2_W/m^2_Z)} \label{eqtwenty}
\ee

\noi where $\Delta r$ represents finite radiative corrections. Similarly, one finds \cite{marcssi93}

\beq
G_\mu (1-\Delta \hat r) & = & \frac{2\sqrt{2}\pi\alpha}{m^2_Z\sin^22\theta_W\mzms} \label{eqtwentyone} \\
G_\mu(1-\Delta r_{\overline{MS}}) & = & \frac{\pi\alpha}{\sqrt{2}m^2_W\sinsthw\mzms} \label{eqtwentytwo}
\eeq

\noi where $\Delta\hat r$ and $\Delta r_{\overline{MS}}$ represent distinct finite radiative corrections with different sensitivities to $m_H$ and New Physics.

To use those natural relations, one must precisely specify the definitions of $\alpha$, $G_\mu$, $m_Z$, $m_W$ and $\sinsthw\mzms$ and their experimental values . Details are given in the next section.

\section{Renormalized Parameters}

To properly utilize the natural relations in Section~3, requires the calculation of radiative corrections to $\alpha$, $G_\mu$, $m_Z$, $m_W$, $\sinsthw\mzms$ and the reactions from which they are extracted. That in turn assumes well specified definitions of those parameters and precise determinations of their values. This procedure has matured during the past 30 years to a very refined level where even 2 loop corrections have been included \cite{prd69053006}. Let me briefly review the current situation.

\subsection{$\mathbf{\alpha}$}

The fine structure constant $\alpha$, is one of the most precisely measured quantities in physics. It is usually defined at $q^2=0$ via a subtraction of all vacuum polarization effects (infinite and finite). That prescription leads to the usual fine structure constant of atomic physics which is appropriate for long distance phenomena. From the comparison of the electron anomalous magnetic moment theory and experiment, one finds \cite{hepph0402206}

\be
\alpha\equiv \frac{e^2(0)}{4\pi} = \frac{e^2_0}{4\pi(1+\pi(0))} = 1/137.03599890 (50) \label{eqtwentythree}
\ee

Absorbed into that definition are charged lepton, quark, $W^\pm$ etc loop effects that polarize the vacuum and renormalize $\alpha_0$ to the observed $\alpha$. In fact, all charged elementary particles, including bound states, e.g.\ $\pi^+\pi^-$, contribute. However, since all low energy QED experiments depend essentially on the same $\alpha$, those effects are lost in the comparison. If one defines, a high energy alpha

\be
\alpha(k^2)\equiv \frac{\alpha_0}{1+\pi(k^2)} \label{eqtwentyfour}
\ee

\noi then for large $k^2$, the vacuum polarization effects are manifested as logarithmic corrections

\be
\alpha^{-1}(k^2) = \alpha^{-1}+\frac{1}{3\pi} \ell n \frac{m^2_e}{k^2} + \frac{1}{3\pi} \ell n \frac{m^2_\mu}{k^2} + \cdots \label{eqtwentyfive}
\ee

\noi Of course, hadronic vacuum polarization effects are not as easily illustrated. Fortunately, they can be evaluated via a dispersion relation using $e^+e^-$ annihilation data as a function of the cm energy $\sqrt{s}$

\be
R(s) \equiv \frac{\sigma(e^+e^-\to{}{\rm hadrons})}{\sigma(e^+e^-\to\mu^+\mu^-)} \label{eqtwentysix}
\ee

\noi which incorporates long distance (non-perturbative) effects as well as perturbative QCD

\be
\pi(k^2)_{\rm had} -\pi(0)_{\rm had} = \frac{\alpha k^2}{3\pi} \int^\infty_{4m^2_\pi} ds \frac{R(s)}{s(s-k^2)} \label{eqtwentyseven}
\ee

\noi Employing that prescription through 5 flavors of quarks, Davier and H\"ocker \cite{plb435427} found a hadronic percentage shift

\be
\Delta\alpha^{(5)}_h=0.02763(16) \label{eqtwentyeight}
\ee

\noi which implies (when leptonic loop effects are included

\be
\alpha^{-1}(m^2_Z) = 128.933(21) \label{eqtwentynine}
\ee

\noi a significant shift from $\alpha^{-1}\simeq137$. When I first studied the running of $\alpha$ back in 1979 \cite{prd20274} (before $m_t$ was known), I crudely estimated $\alpha^{-1}(m^2_Z)\simeq 128.5 (1.0)$. So, the uncertainty has been reduced by a factor of 50! That hadronic uncertainty ($\pm0.00016$) in $\Delta\alpha^{(5)}_h$ translates \cite{marcssi93} into an error for $\Delta r$, $\Delta \hat r$ and $\Delta r_{\overline{MS}}$ in eqs.~(\ref{eqtwenty})--(\ref{eqtwentytwo}) of about $\pm0.0002$ which is small but non-negligible in the determination of $m_H$. 

I should point out that there is some (small) controversy in the extraction of $\Delta\alpha^{(5)}_h$ from data. More recent studies of $e^+e^-\to{}$hadrons suggest a slightly larger value 

\be
\Delta\alpha^{(5)}_h=0.02767(16) \label{eqthirty}
\ee

\noi while using $\tau\to\nu_\tau+{}$hadrons + isospin corrections leads \cite{epjc27497} to (roughly)

\be
\Delta\alpha^{(5)}_h=0.02782 (16) \qquad \tau{\rm ~data} \label{eqthirtyone}
\ee

\noi As we shall subsequently see, a larger $\Delta\alpha^{(5)}_h$ corresponds to a lighter Higgs mass prediction.

The hadronic uncertainty in $\alpha(m^2_Z)$ due to the current error in $e^+e^-\to{}$hadrons data is also correlated with other important quantities. For example, as we shall see in Section~7, it gives rise to the primary uncertainty in the Standard Model prediction for the muon anomalous magnetic moment, $a_\mu$. There, the discrepancy between $e^+e^-$ and  $\tau$ data is more pronounced. It also affects low energy tests of Standard Model weak neutral current predictions where $\gamma$-$Z$ mixing through hadronic loops can be very important \cite{prd531066,prd27552} (see Section~8). Clearly, it is important to improve determinations of $\sigma(e^+e^-\to{}$hadrons) as much as possible. In that regard, measurements of the radiative return process $e^+e^-\to\gamma +{}$hadrons at KLOE, BaBar and Belle are very well motivated, and should be pushed as far as possible.

A related short-distance coupling, $\alpha\mzms$ can be defined \cite{prd20274,prl46163} by modified minimal subtraction at a scale $\mu=m_Z$. Its definition is quite analogous to $\alpha_s(\mu)$ in QCD and was introduced as a convenient way to compare different gauge couplings in GUTS where a unified definition of couplings is simple and appropriate. The quantities $\alpha(m^2_Z)$ discussed above and $\alpha\mzms$ are simply related \cite{marcssi93}

\be
\alpha^{-1}\mzms = \alpha^{-1}(m^2_Z) -0.982 = 127.951 (21) \label{eqthirtytwo}
\ee

\subsection{$\mathbf{G}_\mathbf{\mu}$ - The Fermi Constant}

The Fermi constant, as determined from the total muon decay rate, is denoted by $G_\mu$. That decay rate is obtained from the inverse of the muon lifetime \cite{plb5921}

\be
\tau_\mu=2.197035 (40) \times 10^{-6} {\rm ~sec} \label{eqthirtythree}
\ee

\noi which is already very precisely known and its measurement will be further improved by a factor of 20 in a new PSI experiment. Assuming there are no exotic muon decays \cite{jpgnpp2923} (e.g.\ $\mu\to e+{}$wrong neutrinos) of any appreciable size, the total muon decay rate, $\tau^{-1}_\mu$, is calculable in the Standard Model. It corresponds to a radiative inclusive sum $\Gamma(\mu\to e\nu\bar\nu(\gamma)) = \Gamma(\mu\to e\nu\bar\nu) + \Gamma(\mu\to e\nu\bar\nu(\gamma)) + \Gamma(\mu\to e\nu\bar\nu\gamma\gamma) + \Gamma(\mu\to e\nu\bar\nu e^+e^-)\dots$

Including electroweak Standard Model radiative corrections, one absorbs most loop effects into the definition of a renormalized $G_\mu$ and separates out the others specific to muon decay (calculated in an effective 4 fermion local V-A theory). In that way, one obtains \cite{jpgnpp2923,prl131652}

\be
\tau^{-1}_\mu = \frac{G^2_\mu m^5_\mu}{192\pi^3} f\left(\frac{m^2_e}{m^2_\mu}\right) \left(1+\frac35 \frac{m^2_\mu}{m^2_W}\right) (1+RC) \label{eqthirtyfour}
\ee

\noi where

\be
f(x) = 1-8x + 8x^3 -x^4 -12x^2 \ell n x \label{eqthirtyfive}
\ee

\noi is a phase space factor and the separated radiative corrections are given (to 2 loop order) by

\be
RC=\frac{\alpha}{2\pi} \left(\frac{25}{4} -\pi^2\right) \left(1 + \frac{\alpha}{\pi} \left( \frac23 \ell n \frac{m_\mu}{m_e} -3.7 \right) +\left(\frac{\alpha}{\pi}\right)^2 \left(\frac49 \ell n^2 \frac{m_\mu}{m_e} -2.0 \ell n \frac{m_\mu}{m_e}\right)\right) +\cdots \label{eqthirtysix}
\ee

\noi where leading and next-to-leading 3 loop effects are also included. Comparing eqs.~(\ref{eqthirtythree})--(\ref{eqthirtysix}), one finds

\be
G_\mu = 1.16637 (1) \times 10^{-5} {\rm ~GeV}^{-2} \label{eqthirtyseven}
\ee

\noi which makes the Fermi constant the most precisely determined electroweak parameter. (I do not consider $\alpha$ as electroweak.)

What types of other radiative corrections beyond eq.~(\ref{eqthirtysix}) have been absorbed into $G_\mu$. There are many vertex, self-energy and box diagrams that are effectively in $G_\mu$. However, the most interesting are those that contribute to the $W$ propagator self-energy that go into the $W$ boson mass and wavefunction renormalization. Included in that category are 1) a top-bottom loop \cite{prd22471,npb12389}, 2) a Higgs loop contribution to the $W$ self-energy \cite{npb84132} and 3) Potential New Physics loops from, for example, as yet unknown, very heavy fermion loops.

The information in $G_\mu$ is extremely valuable; but, can it be retrieved? If we compare $G_\mu$ obtained from the muon lifetime with say the tau partial decay rate determination of $G_F$ via $\Gamma(\tau\to e\nu\bar\nu (\gamma))$, most loop effects are common to both and cannot be probed. (Tree level differences in $\mu$ and $\tau$ decays due to excited $W^*$ bosons from extra dimensions \cite{prd60093006} or charged Higgs exchange \cite{npb403} in 2 doublet models can be studied in such a comparison.)

The loop information in $G_\mu$ can be exposed by comparing it with $\alpha$, $m_Z$, $m_W$ and $\sinsthw\mzms$ via the natural relations in eqs.~(\ref{eqtwenty})--(\ref{eqtwentytwo}). It is embodied in the radiative corrections. So, for example $\Delta r$ obtained by compaing $\alpha$, $G_\mu$, $m_Z$ and $m_W$ will depend on $m_t$, $m_H$ and any heavy new particle contributions to $W$ propagator loops. The usual approach in that comparison is to start by ignoring the possibility of New Physics and use $\Delta r$ to extract information regarding $m_t$ and $m_H$. However, now that $m_t$ is fairly well determined (after the $D\emptyset$ update) from direct measurements

\be
m_t = 178.0 \pm 4.3 {\rm ~GeV} \label{eqthirtyeight}
\ee

\noi one can use $\Delta r$ to focus on $m_H$ alone. That procedure will be applied in Section~5.

\subsection{$\mathbf{m}_{\mathbf{Z}}$ and $\mathbf{m}_{\mathbf{W}}$}

The $W^\pm$ and $Z$ masses are also ingredients in the natural relations among masses and couplings. Because the $W^\pm$ and $Z$ bosons are unstable particles, there is some ambiguity in the definition of their masses. Masses are often defined as the real part of the propagator pole $m^2= Re s_0$. The width is then derived from the imaginary part of the pole. Those definitions are gauge independent. In the case of the $W$ and $Z$ bosons a slightly different (also gauge independent) mass definition is used (by convention) \cite{plb259373}

\beq
m^2_Z & = & m^2_Z (\rm real~part~of~pole) + \Gamma^2_Z \label{eqthirtynine} \\
m^2_W & = & m^2_W (\rm real~part~of~pole) + \Gamma^2_W \label{eqforty}
\eeq

\noi (The width contributions are relatively important.) With those definitions, one uniquely specifies the radiative corrections in $\Delta r$. They are actually calculated from the $Z$ and $W$ self-energy diagrams which (as in the case of $G_\mu$) depend on $m_t$, $m_H$ and potential New Physics.

Experimentally the LEPI (very precise) $Z$ pole measurements found (using the definition in eq.~(\ref{eqthirtynine}))

\be
m_Z = 91.1875 (21) {\rm ~GeV} \label{eqfortyone}
\ee

\noi More recently, the $W$ mass was determined independently at LEPII and the Tevatron CDF and $D\emptyset$ experiments (using the definition in eq.~(\ref{eqforty}))

\be
\begin{array}{ll}
m_W = 80.412 (42) {\rm ~GeV} \qquad & {\rm LEPII} \\
m_W = 80.454 (59) {\rm ~GeV} \qquad & {\rm Tevatron} \end{array}
\label{eqfortytwo} 
\ee

\noi Those values are consistent and average to

\be
m^{\rm ave}_W = 80.426 (34) {\rm ~GeV} \label{eqfortythree}
\ee

\noi I will subsequently use that average.

\subsection{$\mathbf{\sinsthw\mzms}$}

The final quantity needed in the natural relationships is a renormalized weak mixing angle $\theta^R_W$. It is related to $\sinsthzw$ via

\be
\sin^2\theta^R_W \equiv \sinsthzw + \delta s^2 \label{eqfortyfour}
\ee

\noi where $\delta s^2$ is an infinite counterterm plus possible finite parts that depend on the specific $\theta^R_W$ definition employed.

There are many ways to define the renormalized weak mixing angle. An experimental favorite is to define an effective angle, $\sin^2\theta^{\rm eff}_W$ via the $Z\to \mu^+\mu^-$ forward-backward asymmetry at the $Z$ pole, i.e.\ absorb radiative corrections to that specific process in the definition. Although popular, I find that approach very awkward, since the details sit in someone's computer codes. Instead, I prefer to define $\sinsthw(\mu)_{\overline{MS}}$ (modified minimal subtraction) via \cite{marcssi93,prl46163}

\be
\sinsthw(\mu)_{\overline{MS}} \equiv \frac{e^2(\mu)_{\overline{MS}}}{g^2_2(\mu)_{\overline{MS}}} \label{eqfortyfive}
\ee

\noi That unphysical definition is particularly convenient for GUTS as well as most new calculations, since the $\overline{MS}$ prescription is easily applied (just subtract poles and their related terms).

\be
\delta s^2 = {\rm cons}\cdot \left[ \frac{1}{n-4} + \frac{\gamma}{2} -\ell n \sqrt{4\pi} \right] \label{eqfortysix}
\ee

\noi Actually, the $\overline{MS}$ definition and $\sin^2\theta^{\rm eff}_W$ used at LEP and SLC are numerically very similar for \cite{prd49r1160} $\mu=m_Z$

\be
\sin^2\theta^{\rm eff}_W = \sinsthw\mzms + 0.00028 \label{eqfortyseven}
\ee

\noi In fact, a sensible renormalization approach is to employ an $\overline{MS}$ subtraction and then use eq.~(\ref{eqfortyseven}) to translate to $\sin^2\theta^{\rm eff}_W$ (if desired).

Currently, the $A_{LR}$ asymmetry and leptonic forward-backward asymmetries at the $Z$ pole give

\be
\sinsthw(m_Z)^{\rm leptonic} = 0.23085 (21) \label{eqfortyeight}
\ee

\noi while the forward-backward hadronic $Z$ pole asymmetries (particularly $Z\to b\bar b$) lead to

\be
\sinsthw(m_Z)^{\rm hadronic} = 0.2320 (3) \label{eqfortynine}
\ee

\noi They differ by about 3.5 sigma. As stated in Section~1, I choose to employ the leptonic result in eq.~(\ref{eqfortyeight}) and disregard eq.~(\ref{eqfortynine}) in my subsequent discussion.

\section{Electroweak Radiative Corrections}

A number of precision electroweak measurements have reached the $\pm0.1\%$ level or better. In table~\ref{tabone}, I summarize some of those quantities

\begin{center}
\begin{table}[ht]
\caption{Values of some precisely determined electroweak parameters \label{tabone}}
\begin{tabular}{l}
\\
$\alpha^{-1} = 137.03599890 (50)$ \\[4pt]
$G_\mu = 1.16637 (1) \times10^{-5} {\rm ~GeV}^{-2}$ \\[4pt]
$m_Z = 91.1875 (21)$ GeV \\[4pt]
$m_W = 80.426 (34)$ GeV \\[4pt]
$\sinslep = 0.23085 (21)$ \\[4pt]
$\Gamma_Z = 2.4952 (23)$ GeV \\[4pt]
$\Gamma(Z\to \ell^+\ell^-) = 83.984 (86)$ MeV \\[4pt]
$\Gamma(Z\to{}{\rm invisible}) = 499.0 (1.5)$ MeV
\end{tabular}
\end{table}
\end{center}

\noi Because the electroweak corrections to those quantities have been computed and are connected by natural relations, they provide powerful constraints on $m_H$ and New Physics effects. Although I will not discuss the $Z$ width properties, they are competitive with the other measurements in table~\ref{tabone} when it comes to certain types of New Physics.

One of the original utilizations of radiative corrections and precision measurements was to bound the top quark mass before the top quark discovery. Those studies gave the bound $m_t<200$ GeV and favored a value around 165 GeV\null. Later, the top quark was discovered at Fermilab and its mass settled down at $174.3\pm5.1$ GeV\null. More recently, a new $D\emptyset$ analysis suggests $m_t\simeq180$ GeV and the average top mass is now

\be
m_t ({\rm pole}) = 178.0\pm4.3 {\rm ~GeV} \label{eqfifty}
\ee

\noi By the way, that pole mass can be related to an $\overline{MS}$ defined mass \cite{prl622793}

\be
m_t(m_t)_{\overline{MS}} = m_t({\rm pole}) \left[ 1-\frac43\, \frac{\alpha_s(m_t)}{\pi} + \cdots \right] \label{eqfiftyone}
\ee

\noi which is about (5\%) 9 GeV smaller. The $\overline{MS}$ mass is often more appropriate for radiative corrections calculations.

The natural relations among the quantities in table~\ref{tabone} are very sensitive to $m_t$ and some types of new physics. They are much less dependent on $m_H$. For example, the $\Delta r$ in eq.~(\ref{eqtwenty}) has the following $m_t$ and $m_H$ dependence \cite{prd22471}

\beq
& \Delta r (m_t, m_H) = 1-\frac{\pi\alpha}{\sqrt{2} G_\mu m^2_W (1-m^2_W/m^2_Z)} &  \nonumber \\
& \Delta r\simeq \frac{\alpha}{\pi s^2} \left\{ -\frac{3}{16}\, \frac{m^2_t}{m^2_W} \, \frac{c^2}{s^2} + \frac{11}{48} \ell n \frac{m^2_H}{m^2_Z} \right\} +0.070 + 2 {\rm loops} & \label{eqfiftytwo} \\
& s^2 = \sinsthw \quad, \quad c^2 = \cos^2\theta_W & \nonumber 
\eeq

\noi where the $0.070$ contribution comes mainly from the same vacuum polarization which shift $\alpha \simeq 1/137$ to $\alpha\mzms \simeq 1/128$, approximately a 7\% effect. Similar types of corrections occur for $\Delta \hat r$

\be
\Delta \hat r (m_t, m_H) = 1 - \frac{2\sqrt{2}\pi\alpha}{G_\mu m^2_Z \sin^22\theta_W\mzms} \label{eqfiftythree}
\ee

\noi although it is somewhat less sensitive to $m_t$ and $m_H$. On the other hand, the radiative correction derived from eq.~(\ref{eqtwentytwo})

\be
\Delta r_{\overline{MS}} = 1- \frac{\pi\alpha}{\sqrt{2}m^2_W G_\mu\sinsthw\mzms} \label{eqfiftyfour}
\ee

\noi includes the 0.07 but has almost no dependence on $m_t$ or $m_H$. Fo that reason, $\Delta r_{\overline{MS}}$ provides a consistency check on the Standard Model and a more direct probe for new physics. It is predicted to be

\be
\Delta r_{\overline{MS}} = 0.0695 (5) \label{eqfiftyfive}
\ee

\noi where the uncertainty corresponds to a generous range in $m_t$ and $m_H$ and I have used (see eq.~(\ref{eqthirty}))

\be
\Delta \alpha^{(5)}_h = 0.02767 (16) \label{eqfiftysix}
\ee

\noi If eq.~(\ref{eqfiftyfour}) is found to disagree with eq.~(\ref{eqfiftyfive}), it would indicate new physics or (perhaps more likely) a mistake in the input.

Let me check the consistancy of $m_W$ and $\sinslep$ in table~\ref{tabone}. Inserting those values in eq.~(\ref{eqfiftyfour}) gives

\be
\Delta r_{\overline{MS}} = 0.0692 (11) {\rm ~~for~~} \sinslep = 0.23085 (21) \label{eqfiftyseven}
\ee

\noi which is in very good accord with eq.~(\ref{eqfiftyfive}). On the other hand, employing $\sinshad = 0.2320 (3)$ in that relation leads to

\be
\Delta r_{\overline{MS}} = 0.0738 (14) {\rm ~~for~~} \sinsthw^{\rm hadronic} = 0.2320 (3) \label{eqfiftyeight}
\ee

\noi which is inconsistent with eq.~(\ref{eqfiftyfive}) at about the $3\sigma$ level. That discrepancy illustrates why I rejected $\sinshad$ for being inconsistent with $m_W$ in the Standard Model. They can be rendered consistent only if new physics is introduced.

A convenient set of formulas that nicely illustrate the relationshsip between $m_W$ and $\sinsthw\mzms$ and various input parameters has been given by Ferroglia, Ossola, Passera and Sirlin \cite{prd65113002} (to one and partial two loop order). Normalized to my input

\beq
m_W/({\rm GeV}) & = & 80.409 -0.507 \left(\frac{\Delta \alpha^{(5)}_h}{0.02767} -1 \right) + 0.542 \left[\left(\frac{m_t}{178 {\rm GeV}}\right)^2 -1 \right] \nonumber \\
& & -0.05719 \ell n (m_H/100 {\rm ~GeV}) - 0.00898 \ell n^2 (m_H/100 {\rm ~GeV}) \label{eqfiftynine} \\
\sinsthw\mzms & = & 0.23101 + 0.00969 \left(\frac{\Delta\alpha^{(5)}_h}{0.02767} -1\right) -0.00277 \left[ \left(\frac{m_t}{178 {\rm ~GeV}}\right)^2-1\right] \nonumber \\
& & +0.0004908 \ell n (m_H/100 {\rm ~GeV}) + 0.0000343 \ell n^2 (m_H/100 {\rm ~GeV}) \label{eqsixty}
\eeq

\noi Those formulas can be inverted to predict $m_H$ for a given $m_W$ or $\sinsthw\mzms$. Their predictions are illustrated in the next section.

\section{Higgs Mass Prediction}

Employing the formulas in eqs.~(\ref{eqfiftynine}) and (\ref{eqsixty}) along with the range of $m_t$ and $\Delta\alpha^{(5)}_h$ in eqs.~(\ref{eqfifty}) and (\ref{eqthirty}), one finds the Higgs mass predictions

\beq
&m_W = 80.426 (34) {\rm ~GeV}\to m_H=74^{+83}_{-47} {\rm ~GeV}, &<238 {\rm ~GeV~(95\%~CL)} \label{eqsixtyone} \\
&\sinsthw\mzms = 0.23085 (21) \to m_H = 71^{+48}_{-32} {\rm ~GeV}, &<167 {\rm ~GeV~(95\%~CL)} \label{eqsixtytwo}
\eeq

\noi Those constraints are very consistent with one another. A full 2 loop analysis \cite{prd69053006,hepph0406334} lowers the Higgs mass prediction in eq.~(\ref{eqsixtyone}) to $62^{+78}_{-43}$ GeV, $<216$ GeV\null. Combining, that result with eq(62) implies 

\be
m_H=68^{+45}_{-30} {\rm ~GeV} \quad , \quad <154 {\rm ~GeV~(95\%~CL)} \label{eqsixtythree}
\ee

\noi Such a low Higgs mass is very suggestive of supersymmetric models in which one expects $m_H\lsim135$ GeV for the lightest supersymmetric scalar.

 A larger $\Delta\alpha^{(5)}_h$ as suggested by $\tau$ decay data lowers the value of $m_H$ further. If one employs $\sinshad=0.2320 (3)$ alone, it leads to $m_H\simeq500^{+350}_{-240}$ GeV, which is inconsistent with eqs.~(\ref{eqsixtyone}) and (\ref{eqsixtytwo}). That result illustrates an interesting feature. Because there is only a logarithmic sensitivity to $m_H$, the uncertainty in $m_H$ scales with its central value. Because the central values in eqs.~(\ref{eqsixtyone}) and (\ref{eqsixtytwo}) are small, the errors are also small. If the central value of $m_H$ were much larger, the errors would scale up and we would likely conclude that there was not much of a constraint on $m_H$.

So, it seems that $m_W$ and $\sinslep$ are very consistent with one another and both are indicating a relatively light Higgs scalar.

\subsection{$\mathbf{S}$ Parameter Constraints}

If new physics in the form of heavy fermion loops contribute to gauge boson self-energies, they will manifest themselves in the natural relations via $\Delta r$, $\Delta \hat r$ and $\Delta r_{\overline{MS}}$. A nice parametrization of such effects has been given by Peskin and Takeuchi \cite{prl65964} in terms of an isospin conserving quantity, $S$, and isospin violating parameter $T$. Full discussions of the sensitivity to $S$ and $T$ via precision measurements are given in ref.\ \cite{marcssi93}. Here, I will mainly comment on $S$.

Bounds on $S$ and $T$ have been given using global fits to all electroweak data. One such recent fit gives \cite{jerlerref4}

\beq
S & \simeq & -0.1\pm0.1 \nonumber \\
T & \simeq & -0.1\pm0.1 \label{eqsixtyfour}
\eeq

\noi which are consistent with zero and imply no evidence for new physics. (In the Standard Model, one expects $S=T=0$, modulo the $m_H
$ uncertainty.) A simple way to constrain $S$ comes from a comparison of $m_W$ and $\sinsthw\mzms$. In fact, there is a very nice, but little known formula \cite{prl652963}

\be
S\simeq 118 \left[ 2 \frac{m_W-80.409 {\rm ~GeV}}{80.409 {\rm ~GeV}} + \frac{\sinsthw\mzms-0.23101}{0.23101} \right] \label{eqsixtyfive}
\ee

\noi Using the values of $m_W$ and $\sinsthw\mzms$ in table~\ref{tabone} gives 

\be
S=-0.03\pm0.1\pm0.1 \label{eqsixtysix}
\ee

\noi which is nearly as constraining as eq.~(\ref{eqsixtyfour}), but more transparent in its origin.

The constraint in eq.~(\ref{eqsixtyfour}) or (\ref{eqsixtysix}) can be used to rule out or limit various new physics scenarios. Each new heavy chiral fermion doublet contributes \cite{prl65964,npb248589} $+1/6\pi$ to $S$. A full 4th generation of quarks and leptons (4 doublets) should contribute $+0.21$ to $S$ and that seems to be ruled out or at least unlikely. It also strongly disfavors dynamical symmetry breaking models which generally have many heavy fermion doublets and tend to give $S\simeq {\cal{O}}(1)$. In fact, the constraint on $S$ is rather devastating for most New Physics scenarios, with the exception of supersymmetry or other symmetry constrained theories where one expects $S\simeq0$. 

If instead of $\sinslep$, we compare $\sinshad = 0.2320 (3)$ with $m_W$, then we find from eq.~(\ref{eqsixtyfour}) $S\simeq0.56 \pm0.18$. At face value, that would seem to suggest the appearance of New Physics in $S$ at the 3 sigma level. However, more likely, it represents a problem with $\sinshad$ from some as yet unidentified systematic effect. It supports my argument for disregarding $\sinshad$.

Clearly, it would be very nice to reduce further the uncertainties in $m_W$ and $\sinsthw\mzms$ as a means of pinpointing $m_H$ and determining $S$ more precisely. Toward that end, a giga $Z$ factory ($\gsim10^9 Z$ bosons) with polarized $e^+$ and $e^-$ beams could potentially measure $\sinslep$ to an incredible $\pm0.00002$! Also, running near the $W^+W^-$ threshold, it could determine $m_W$ to about $\pm0.006$ GeV\null. At those levels, $\Delta m_H/m_H$ could be predicted to $\pm5\%$ or $S$ constrained to $\pm0.02$. Such advances would be spectacular probes of the Standard Model and beyond. (Another means of improving $\sinslep$ will be discussed in Section~8.)

\section{The Muon Anomalous Magnetic Moment}

Currently, there is a 2.7 sigma discrepancy between the experimental and Standard Model (SM)  values of the muon anomalous magnetic moment, $a_\mu$. That difference could be an experimental issue, an incorrect evaluation of hadronic loops or New Physics. In this section, I will review the status of $a^{\rm SM}_\mu$ and argue in favor of the New Physics interpretation. I also discuss an indirect $a_\mu-m_H$ connection \cite{np116437} via $e^+e^-\to{}$hadrons data. Collectively, it appears that precision electroweak data plus $a_\mu$ may be hinting at the presence of supersymmetry \cite{prd64013014}.

Let me start with some early history. A great success of the Dirac equation (1928) \cite{prsa117610} (which married quantum mechanics and special relativity) was its prediction (or postdiction) that the gyromagnetic ratio or $g$ factor of the electron should be 2. Later, in 1948, Schwinger \cite{pr73416L} showed that quantum loops give rise to a deviation in $g_e$ from 2, the $g_e-2$ anomaly. He computed from 1 loop effects

\be
a_e=\frac{g_e-2}{2} = \frac{\alpha}{2\pi} \simeq 0.00116 \label{eqsixtyseven}
\ee

\noi a simple beautiful prediction that was confirmed by experiment and heralded as a great triumph for QED\null. It is now routinely calculated by physics students in primary school.

Following that early success, experiments measured $a_e$ and $a_\mu$ (the muon anomalous magnetic moment) ever more precisely. At the same time, higher order loop effects have been computed. In fact, a nice synergy has existed. As experiments became more precise, they would often disagree with theory. The disagreement would then lead to errors in theory being uncovered or force theorists to compute yet higher order effects. Currently, the state of theory and experiment are both impressive. They are testimonies to the capabilities of theorists and experimentalists when driven by a challenging (stimulating) problem.

In the case of the electron (or positron), $a_e$ has been computed through 4 loop order in QED and small SM electroweak and hadronic loop effects have been evaluated. One finds in total \cite{arnps542004}

\beq
a^{\rm SM}_e & = & \frac{\alpha}{2\pi} -0.328478444 \left(\frac{\alpha}{\pi}\right)^2 + 1.181234 \left(\frac{\alpha}{\pi}\right)^3 - 1.7502 \left(\frac{\alpha}{\pi}\right)^4 \nonumber \\
& & +1.6\times10^{-12} \label{eqsixtyeight}
\eeq

\noi where the last term stems from strong (hadronic 2 loop) and electroweak corrections. They are of order $\left(\frac{\alpha}{\pi}\right)^2 m^2_e/m^2_\rho \simeq 2\times10^{-12}$ and $\frac{\alpha}{\pi} m^2_e/m^2_W \simeq 10^{-13}$ respectively.

The SM prediction in eq.~(\ref{eqsixtyeight}) is to be compared with the (Nobel prize winning) experimental results \cite{plb5921}

\beq
a^{\rm exp}_{e^-} & = & 0.0011596521884 (43) \nonumber \\
a^{\rm exp}_{e^+} & = & 0.0011596521879 (43) \label{eqsixtynine}
\eeq

\noi Comparison of eqs.~(\ref{eqsixtyeight}) and (\ref{eqsixtynine}) currently provides the best determination of $\alpha$ (see eq.\ref{eqtwentythree}). That determination in consistent with other (less precise) condensed matter and atomic physics determinations \cite{rmp72356}. New physics effects (much like strong and electroweak) are expected to contribute $\sim m^2_e/\Lambda^2$ to $a_e$ where $\Lambda$ is the scale of New Physics. So, the electron anomalous magnetic moment is not very sensitive to such effects. Note, a new experiment underway at Harvard aims to further improve $a^{\rm exp}_e$ by a factor of 15.

The muon anomalous magnetic moment is about $m^2_\mu/m^2_e\simeq 40,000$ times more sensitive than $a_e$ to New Physics, as well as hadronic and electroweak loops. The experimental uncertainty in $a^{\rm exp}_\mu$ is less than 100 times worse than $a^{\rm exp}_e$; so, $a^{\rm exp}_\mu$ is clearly a much better place to look for New Physics. Of course, one must also do a much better job of computing strong and electroweak contributions to $a^{\rm SM}_\mu$ because of their relative enhancement for the muon.

The E821 experiment at Brookhaven has completed its measurements of $a^{\rm exp}_{\mu^+}$ and $a^{\rm exp}_{\mu^-}$ (see talk by P. Shagin). They are consistent with one another and average to \cite{bennett}

\be 
a^{\rm exp}_\mu = 116592080 (58) \times 10^{-11}, \label{eqseventy}
\ee

\noi about a factor of 14 improvement over the classic CERN experiments of the 1970s. A new upgraded version of that experiment E969 has been approved, but requires funding. It would reduce the error in eq.~(\ref{eqseventy}) by a factor of 2.5, to about $\pm 23\times10^{-11}$. As we shall see, there are strong reasons to push for such improvement. 

To utilize the result in eq.~(\ref{eqseventy}) requires a Standard Model calculation of comparable precision. That theory prediction is generally divided into 3 parts

\be
a^{\rm SM}_\mu = a^{\rm QED}_\mu + a^{\rm EW}_\mu + a^{\rm Hadronic}_\mu \label{eqseventyone}
\ee

\noi The QED part results from quantum loops involving photons and
leptons. They have been computed through 4 loops and estimated at the 5
loop level. Including the recent update of the 5 loop estimate reported by Kinoshita, one finds \cite{hepph0402206}

\beq
a^{\rm QED}_\mu & = & \frac{\alpha}{2\pi} + 0.765857376
\left(\frac{\alpha}{\pi}\right)^2 + 24.05050898
\left(\frac{\alpha}{\pi}\right)^3 + 131.0
\left(\frac{\alpha}{\pi}\right)^4 \nonumber \\
& & \qquad + 677(40) \left(\frac{\alpha}{\pi}\right)^5 +\cdots
\label{eqseventytwo}
\eeq

\noi Employing the value of $\alpha$ determined \cite{hepph0402206} from
the electron $a_e$

\be
\alpha^{-1} (a_e) = 137.03599890 (1.5) (3.1) (50) \label{eqseventythree}
\ee

\noi leads to

\be
a^{\rm QED}_\mu = 116584719 (1) \times 10^{-11} \label{eqseventyfour}
\ee

\noi That result is somewhat larger \cite{arnps542004} than the generally
quoted value of a few years ago, but now has a much firmer basis with
an insignificant error assigned to it.

The electroweak contribution from $W$ and $Z$ bosons is given at the
one loop level by \cite{jackiw}

\be
a^{\rm EW}_\mu (1{\rm~loop}) = \frac{5}{24}\, \frac{G_\mu}{\sqrt{2}}\,
\frac{m^2_\mu}{\pi^2} \left( 1+ \frac{1}{5}
(1-4\sin^2\theta_W)^2\right) = 194.8\times10^{-11} \label{eqseventyfive}
\ee

\noi Two loop effects turned out to be unexpectedly large
\cite{kukhto,czarone,czartwo} 

\be
a^{\rm EW}_\mu (2{\rm~loop}) = -40.7 (1.0)(1.8) \times 10^{-11}
\label{eqseventysix} 
\ee

\noi and revealed some very interesting features (such as novelties in
hadronic triangle anomaly diagrams
\cite{czarone,czartwo,marcone,peris}). Finally, the 3 loop leading logs
were found to be negligible, ${\cal{O}}(10^{-12})$ \cite{czartwo,degrassi}. In total, one finds

\be
a^{\rm EW}_\mu = 154 (1) (2) \times 10^{-11} \label{eqseventyseven}
\ee

\noi where the first error stems from hadronic triangle diagram
uncertainties and the second is primarily due to the Higgs mass
uncertainty \cite{czartwo}. 

The final computed effect due to hadronic loops manifests itself at
${\cal{O}}\left(\left(\frac{\alpha}{\pi}\right)^2\right)$ from
hadronic vacuum polarization loops (along with the ${\cal{O}}
\left(\left(\frac{\alpha}{\pi}\right)^3\right)$ photonic corrections to the
hadronic vacuum polarization). There are also additional ${\cal{O}}
\left(\left(\frac{\alpha}{\pi}\right)^3\right)$ hadronic loops,
including the infamous hadronic light by light loops. Including the
recent KLOE data for $e^+e^-\to{}$hadrons${}+\gamma$ leads to \cite{hoecker}

\be
a^{\rm Hadronic}_\mu (\rm vacuum~pol) = 6934 (53) (35)_{\rm RC} \times
10^{-11} \label{eqseventyeight}
\ee

\noi where RC stands for uncertainties in the radiative corrections to
$e^+e^-\to{}$hadrons data. The additional 3 loop hadronic contributions
were found to be \cite{arnps542004}

\be
a^{\rm Hadronic}_\mu (3{\rm~loop}) = 22 (35) \times 10^{-11}
\label{eqseventynine} 
\ee

\noi with the error dominated by hadronic light by light uncertainties.

Adding Eqs. (\ref{eqseventyfour}), (\ref{eqseventyseven}), (\ref{eqseventyeight}) and
(\ref{eqseventynine}) leads to the Standard Model prediction

\be
a^{\rm SM}_\mu  =  116591829 (53) (35)_{\rm RC} (35)_{\rm LBL} (3) \times
10^{-11} \label{eqeighty} 
\ee

\noi The overall uncertainty in that prediction is fairly well matched
to the current experimental uncertainty in Eq.~(\ref{eqseventy}) and
leads to the 2.7 sigma discrepancy for $a^{\rm exp}_\mu - a^{\rm
SM}_\mu$

\be
\Delta a_\mu  =  a^{\rm exp}_\mu - a^{\rm SM}_\mu = 251(93)\times10^{-11} \label{neweqeightyone}
\ee

\noi It is anticipated \cite{arnps542004} that improved measurements of
$e^+e^-\to{}$hadrons$+\gamma$ at KLOE and BaBar will (relatively soon)
reduce the error in Eq. (\ref{eqseventyeight}) by about a factor of 2. More
problematic than the error in Eq.~(\ref{eqeighty}) at this time is a
discrepancy between $e^+e^-\to{}$hadrons in the $I=1$ channel and
$\tau\to\nu_\tau+{}$hadrons data, even after isospin violating
corrections are taken into account. Indeed, using
$\tau^-\to\nu_\tau\pi^-\pi^0$ data \cite{epjc27497} around (and above) the
rho resonance 
in the hadronic vacuum polarization dispersion relation increases
$a^{\rm Hadronic}_\mu$ by about $+137\times 10^{-11}$. Such a shift
would reduce the $a_\mu$ discrepancy to a not so interesting 1.3 sigma effect

\be
a^{\rm exp}_\mu - a^{\rm SM}_\mu = 114 (89) \times 10^{-11} \qquad
(\tau^-\to\nu_\tau \pi^-\pi^0 {\rm~data}) \label{eqeightyone}
\ee

\noi However, it appears that not all isospin violating corrections to
the tau data have been applied. In particular, corrections due to the
$\rho^-$-$\rho^0$ mass and width differences are not reliably known and
hence have not been fully applied \cite{hoecker}. So, for now, the tau data is
provocative but not on as firm a footing as $e^+e^-$ annihilation data
which is more reliably applied in the dispersion relation. Hopefully,
ongoing studies of isospin violating effects will eventually resolve
the $\tau$-$e^+e^-$ disagreement.

The total uncertainty in Eq.~(\ref{eqeighty}) (errors added in quadrature)
is about $\pm73\times10^{-11}$. How much (further) can that be reduced?
Improvements in $e^+e^-\to{}$hadrons data should reduce the hadronic
vacuum polarization contribution uncertainty to about
$\pm30\times10^{-11}$. That would leave the hadronic light by light
uncertainty, $\pm35\times10^{-11}$ as the dominant error. But that
assigned error  is rather conservative and can probably be reduced
to about $\pm20\times10^{-11}$ by refining the approach of Melnikov and
Vainshtein \cite{melnikov}. That would lead to a total uncertainty of
$\pm36\times10^{-11}$ in $a^{\rm SM}_\mu$, which is well matched to the
goal of E969 ($\pm23\times10^{-11}$) for $a^{\rm exp}_\mu$. Should
those improvements in theory and experiment be achieved and the
discrepancy central value remain approximately unchanged, it would be
elevated to about a 6 sigma effect, which would certainly be
interpreted as a sign of ``New Physics''. Reducing the uncertainty in
$a^{\rm SM}_\mu$ much below $\pm36\times10^{-11}$ currently appears to
be very difficult, but who knows.

The  discrepancy in Eq.~(\ref{neweqeightyone}), $\Delta a_\mu\simeq
251\times 10^{-11}$, is very large. Recall that $a^{\rm
EW}_\mu=154\times10^{-11}$. Can ``New Physics'' give a contribution
larger than $W$ and $Z$ bosons? That might seem unlikely, since
precision studies of $W$ and $Z$ bosons have confirmed the Standard
Model at the $\pm0.1$\%. However, it does not take much of a stretch of
one's imagination to come up with viable explanations.

The leading ``New Physics'' explanation for the discrepancy in
Eq.~(\ref{neweqeightyone}) is supersymmetry \cite{yuan,kosower}.  It enters at the
one loop level via charginos, sneutrinos, neutralinos and sleptons.
Generically, one might expect the SUSY contribution to $a_\mu$ to be
roughly the magnitude of the EW effect in Eq.~(\ref{eqseventyfive}) times
$(m_W/m_{\rm 
SUSY})^2$. The exact prediction is of course model dependent. One can
get a good feel for $a^{\rm SUSY}_\mu$ by taking all SUSY loop masses
to be degenerate and given by $m_{\rm SUSY}$. In that way \cite{moroi},
one finds to 
leading order in large $\tan\beta$ (including 2 loop leading QED log
corrections) \cite{prd64013014} 

\begin{equation}
a^{\rm SUSY}_\mu \simeq ({\rm sign} \mu)\times130\times10^{-11}
\tan\beta \left(\frac{100{\rm~GeV}}{m_{\rm SUSY}}\right)^2
\label{eqeightytwo} 
\end{equation}

\noi where sign$\mu=+$ or $-$ (depending on the sign of the 2 Higgs
mixing term in the Lagrangian) and 

\begin{equation}
\tan\beta = \frac{\langle \phi_2\rangle}{\langle \phi_1\rangle} =
\frac{v_{2}}{v_{1}} \label{eqeightythree}
\end{equation}

\noi is the ratio of Higgs doublet vacuum expectation values.

A significant development, over the last 20 years, has been a change in
the mindset $\tan\beta\simeq1$ to the more likely higher values

\begin{equation}
\tan\beta \simeq 3-40 \label{eqeightyfour}
\end{equation}

\noi which would imply an enhancement of $a^{\rm SUSY}_\mu$. Such an
enhancement characterizes loop induced chiral changing amplitudes
$a_\mu$, electric dipole moments, $\mu\to e\gamma$, $b\to s\gamma$ etc,
rendering them sensitive probes of $\tan\beta$ and supersymmetry.

Equating Eqs. (\ref{neweqeightyone}) and (\ref{eqeightytwo}) leads to the
constraint 

\begin{eqnarray}
{\rm sign} \mu & = & + \label{eqeightyfive} \\
m_{\rm SUSY} & \simeq & 72\sqrt{\tan\beta} {\rm ~GeV} \label{eqeightysix} 
\end{eqnarray}

\noi Those generic implications are very powerful. The first one
eliminates about half of all SUSY models (those with sign$\mu=-$) and
is consistent with $b\to s\gamma$ results. The second (rough) constraint in 
Eq.~(\ref{eqeightysix})
suggests $m_{\rm SUSY} \simeq 100$--500 GeV, just where many advocates
expect it.

If $\Delta a_\mu$ is suggestive of SUSY, it would join other potential
early signs of supersymmetry: 1) SUSY GUT Unification, 2) Precision
measurements that suggest a relatively light Higgs and 3) Dark Matter.
Interestingly, sign$\mu=+$ makes it more likely that underground
detectors will be able to detect dark matter recoil signals, an
exciting possibility.

Are there other viable ``New Physics'' explanations of $\Delta a_\mu$
besides SUSY? Many have been explored. Here, I mention another generic
possibility, radiative muon mass models \cite{prd64013014,marctwo}. In
theories where the bare muon mass $m^0_\mu=0$ and the actual mass is
loop induced, there is a simple relationship between the muon mass and
$a^{\rm New}_\mu$ arising from similar loop effects, such
that \cite{prd64013014,marctwo} 

\be
a^{\rm New}_\mu = {\cal{O}}(1) \times \frac{m^2_\mu}{\Lambda^2}
\label{eqeightyseven}
\ee

\noi where $\Lambda$ is the scale of the underlying new physics
responsible for mass generation. The current deviation would suggest

\begin{equation}
\Lambda \simeq 2{\rm ~TeV} \label{eqeightyeight}
\end{equation}

\noi Examples of such scenarios include: Extra Dimensions, Multi-Higgs,
New Dynamics, SUSY etc.

Is $\Delta a_\mu\simeq 251\times10^{-11}$, as currently suggested by
theory and experiment, a harbinger of supersymmetry or some other ``New
Physics''? If it is SUSY, it implies happy days for the LHC, Dark Matter
Searches, Flavor Changing Neutral Currents (e.g. $\mu\to e\gamma$, $b\to
s\gamma\dots$). Of course, that hint also cries out for further
improvements in $a^{\rm exp}_\mu$ and $a^{\rm SM}_\mu$. Fortunately,
improvement in $a^{\rm SM}_\mu$ by about a factor of 2 from
$e^+e^-\to{}$hadrons${}+\gamma$ data seems likely and an improvement in
$a^{\rm exp}_\mu$ by a factor of 2.5 has been approved (pending
funding). If a deviation of 5--6 sigma in $\Delta a_\mu$ results, it
will be very complementary to more direct searches for ``New Physics''
at the LHC\null. For example, if SUSY particles are discovered at the
LHC, $a^{\rm SUSY}_\mu$ may provide the best determination of
$\tan\beta$, an otherwise difficult parameter to measure.

Experiments such as the muon anomalous magnetic moment challenge our
technical capabilities, computational talents and model building
imaginations. They should be pushed as far as possible.

\subsection{The $\mathbf{a}_{\mathbf{\mu}}$-$\mathbf{m}_{\mathbf{H}}$ Connection}

The difference between experiment and theory in the case of the anomalous magnetic moment of the muon $\Delta a_\mu=251(93)\times10^{-11}$ is suggestive of a supersymmetric loop contribution. However, that interpretation is clouded by a substantial discrepancy between $e^+e^-\to{}$hadrons data used to obtain $a^{\rm Hadronic}_\mu$ and $\tau\to\nu_\tau+{}$hadrons data which should be related to it, up to isospin violating corrections. The tau data is either a red flag, alerting us to an error in $e^+e^-\to{}$hadrons studies or a red herring, misleading us into doubt about $a^{\rm Hadronic}_\mu$.

The $e^+e^-$ and $\tau$ data disagree somewhat near the rho peak and more significantly in its high energy tail $\sqrt{s}\gsim1$ GeV\null. Part of the difference could be pointing to isospin violating effects which give $m_{\rho^\pm}\ne m_{\rho^0}$ and $\Gamma_{\rho^\pm}\ne \Gamma_{\rho^0}$. Indeed, the data seems to suggest \cite{hoecker} $m_{\rho^\pm}-m_{\rho^0}\simeq 2$--3 MeV and about a 2\% broadening of $\rho^0$ relative to $\rho^\pm$. It is harder to explain the difference in the higher energy region.

Since the mass prediction, $m_H$, also depends on $e^+e^-\to{}$hadrons data via $\Delta\alpha^{(5)}_h$ which comes from a dispersion relation, it is interesting to ask what effect the tau data taken at face vaule, has on $m_H$? In the case of $\Delta a_\mu$, tau data helps reduce the discrepancy, but what does it do to the $m_H$ predictions in eqs.~(\ref{eqsixtyone})--(\ref{eqsixtythree}) which already prefer a very light Higgs, below the direct search bounds, but allow some room, in a region suggested by supersymmetry. It turns out that tau data increases $\Delta\alpha^{(5)}_h$ and leads to a further decrease in the predicted value of $m_H$. From eqs.~(\ref{eqfiftynine}) and (\ref{eqsixty}), one finds (roughly)

\beq
m_H & \simeq & 74^{+83}_{-47} {\rm ~exp}\left[ -9\left(\frac{\Delta\alpha^{(5)}_h}{0.02767} -1\right)\right] {\rm ~GeV~from~} m_W \label{eqeightynine} \\
m_H & \simeq & 71^{+48}_{-32} {\rm ~exp}\left[-20 \left(\frac{\Delta\alpha^{(5)}_h}{0.02767} -1\right)\right] {\rm ~GeV~from~} \sinslep \label{eqninety}
\eeq

\noi The prediction from $\sinsthw\mzms$ is much more sensitive to changes in $\Delta\alpha^{(5)}_h$. If one uses $\Delta\alpha^{(5)}_h \simeq 0.02782 (16)$ as suggested by $\tau\to\nu_\tau +{}$hadrons data, it lowers the prediction for $m_H$ in eq.~(\ref{eqsixtytwo}) to $m_H\simeq 64^{+44}_{-30}$ GeV, $<152$ GeV (95\% CL). Although some experimentally allowed region remains, it is suggestive of a near conflict. Indeed, increasing $e^+e^-\to{}$hadrons cross-sections much more, in an effort to completely eliminate the $\Delta a_\mu$ discrepancy, would likely bring the $m_H$ prediction from $\sinslep$ below the 114 GeV experimental lower bound.

So, $\Delta a_\mu$ and $m_H$ together are rather consistent with $e^+e^-\to{}$hadrons data. They suggest that supersymmetry may be the cause of $\Delta a_\mu$ and may be the mechanism responsible for a relatively light (but not too light) Higgs that precision electroweak data seems to favor.

\section{Other Precision Studies}

The $Z$ pole measurements at LEP and the SLC set a high standard for precision, attaining $\pm0.1\%$ (or better) determinations of many electroweak quantities. A similar level of precision has also been achieved in low energy charged current interaction studies: $\mu$, $\tau$, $\pi$, $\beta\dots$decays. In the case of weak neutral current studies at $q^2<< m^2_Z$, experiments have been less precise, only achieving about $\pm1\%$ accuracy, but have nevertheless played an extremely important role in testing the structure of the Standard Model and probing for new physics. An early example is the famous SLAC polarized $eD$ experiment \cite{pl77b347} that measured $A_{LR}$ and established the correctness of the Standard Model's weak neutral current. That experiment set a historical milestone and provided a relatively precise (for its day) measurement of $\sinsthw$.

Atomic parity violation (APV) experiments started out missing the predicted Standard Model effects. Those efforts rebounded with some beautiful measurements, achieving $\pm1\%$ precision in $Cs$ studies \cite{prl61310}. That level of accuracy has played a significant role in ruling out new physics scenarios, via the $S$ parameter \cite{prl652963}. In addition, APV is very sensitive to $Z^\prime$ bosons \cite{prl652963}, leptoquarks, extra dimensions etc.

More recently, deep-inelastic $\nu_\mu N$-scattering has caused some fuss. By measuring $R_\nu\equiv \sigma (\nu_\mu N\to\nu_\mu X)/\sigma (\nu_\mu N\to \mu^-X)$ and $R_{\bar\nu}$, the NuTeV collaboration \cite{prl88091802} at Fermilab found a 3 sigma deviation from Standard Model expectations. That anomaly has called into question aspects of $s\bar s$ and isospin asymmetries in quark distributions and the application of radiative corrections \cite{npb189442} to the data. An alternate explanation could be a very heavy Higgs mass loop effects, but that interpretation conflicts with $m_W$ and $Z$ pole asymmetry results. It will be interesting to see how this deviation ultimately plays out.

\subsection{SLAC E158: Polarized $\mathbf{e}^{\mathbf-}\mathbf{e}^{\mathbf-}$ Asymmetry}

I would like to focus in the remainder of this lecture on a recently completed experiment \cite{kolossi04} at SLAC, E158. That experiment measured the parity violating left-right asymmetry in polarized $e^-e^-$ M\o ller scattering.

\be
A_{LR} = \frac{\sigma (e^-_Le^-\to e^-e^-) - \sigma(e^-_Re^-\to e^-e^-)}{\sigma(e^-_Le^-\to e^-e^-) + \sigma(e^-_Re^-\to e^-e^-)} \label{eqninetytwo}
\ee

\noi It used a polarized 50 GeV $e^-$ beam on a fixed target with $Q^2\simeq 0.026$ GeV$^2$. Experimental details have been given  in the talk by Y. Kolominsky. Here, I will concentrate on some theoretical aspects \cite{marcssi93,prd531066,ap121147,ijmpa132235}.

The $e^-e^-$ asymmetry in eq.~(\ref{eqninetytwo}) is due at tree level to the interference of $\gamma$ and $Z$ exchange amplitudes. In lowest order, it is predicted to be

\be
A_{LR}(e^-e^-) = \frac{G^0_\mu s}{\sqrt{2}\pi\alpha}\, \frac{y(1-y)}{1+y^4+(1-y)^4} (1-4\sinsthzw) \label{eqninetythree}
\ee

\noi where $Q^2=ys$ and $0\le y\le 1$ with

\be
y=\frac{1-\cos\theta_{cm}}{2} \label{eqninetyfour}
\ee

\noi in the cms. Several features of $A_{LR}$ are interesting. It is very small because $G_\mu s\simeq5\times10^{-7}$. In fact, even for $y=1/2$, $A_{LR}$ is expected to be $\sim3\times10^{-7}$ in lowest order. To accurately measure such a small asymmetry with good precision requires about $10^{16}$ events. Those gigantic statistics are only possible because, for fixed target scattering the effective luminosity ($\sim4\times10^{38}cm^{-2}/s$) and cross-section are both enormous. Note also, that there is a $1-4\sinsthzw$ suppression factor which makes $A_{LR}$ very sensitive to the exact value of the weak mixing angle. A feature that follows from that sensitivity implies that the closer $\sinsthw$ is to 1/4, the less important the polarization uncertainty becomes. That property can be very important if one wishes to push the $A_{LR}$ effort to a very high precision determination of the weak mixing angle, since it is very difficult to measure the polarization to much better than $\pm0.4\%$.

Electroweak radiative corrections to $A_{LR}$ are large and important. Roughly speaking, the one loop radiative corrections replace $G^0_\mu(1-4\sinsthzw)$ in eq.~(\ref{eqninetythree}) by \cite{prd531066}

\beq
&& \rho G_\mu \left(1-4\kappa (Q^2)\sinsthw\mzms + \frac{\alpha(m_Z)_{\overline{MS}}}{4\pi s^2} \right. \nonumber \\
&&\left. - \frac{3\alpha\mzms}{32\pi s^2c^2} (1-4s^2) (1+(1-4s^2)^2) + F_1(Q^2)\right) \label{eqninetyfive}
\eeq

\noi where $F_1(Q^2\simeq0)\simeq-0.004(10)$ and $\rho\simeq1.001$. The most  important effect of radiative corrections is in $\kappa(Q^2)$. The quantity $\kappa(Q^2)\sinsthw\mzms$ corresponds to a running $\sinsthw(Q^2)$ which exhibits an interesting $Q^2$ dependence. For $Q^2\simeq0$, as is appropriate for E158, one finds

\beq
\kappa(0) & = & 1-\frac{\alpha}{2\pi s^2} \left\{ \frac13 \sum_f(T_{3f}Q_f-2s^2Q^2_f) \ell n \frac{m^2_f}{m^2_Z} \right. \nonumber \\
& & \left. -\left(\frac72 c^2+\frac{1}{12}\right) \ell n c^2 + \left(\frac79- \frac{s^2}{3}\right) \right\} \simeq 1.0301(25) \label{eqninetysix}
\eeq

\noi where the sum is over all fermions, quarks and leptons. In reality $e^+e^-\to{}$hadrons data must be used in a dispersion relation to evaluate the hadronic part of eq.~(\ref{eqninetysix}).

The 3\% increase in $\sinsthw(Q^2)$, as $Q^2$ ranges from $m^2_Z$ to 0, is very important. It reduces the predicted $A_{LR}$ by about 40\%, a significant reduction. As mentioned before, that reduction also lessens the error from polarization uncertainty in the value of $\sinsthw(0)$ extracted. It would be very interesting to precisely measure the running \cite{ijmpa132235} of $\sinsthw(Q^2)$ for a variety of $Q^2$, to verify the predicted Standard Model behavior.

After taking electroweak radiative corrections into account, E158 has reported a preliminary result \cite{kolossi04}

\be
\sinsthw\mzms=0.2330(11)_{\rm stat} (10)_{\rm syst} \qquad {\rm Preliminary} \label{eqninetyseven}
\ee

\noi (That corresponds to $\sinsthw(0)\simeq0.240$.) The central value is a little high compared to $\sinslep\simeq0.23085(21)$ obtained from the $Z$ pole measurements but is consistent with running. The error is, however, still relatively large. Of course, the main reason for studying $A_{LR}$ off the $Z$ pole is not necessarily to measure $\sinsthw\mzms$. Instead, it is to look for New Physics effects that might be very small when the $Z$ is on mass shell, but more important off resonance. Examples are \cite{prd531066} a $Z^\prime$ boson (e.g.\ the $Z_\chi$ of $S0(10)$), effects of extra dimensions, a potential doubly charge $H^{--}$, electron anapole moments effects, contact interactions etc. For example, a $Z_\chi$ would lead to an increase in $A_{LR}$ by a factor of $1+7 m^2_Z/m^2_{Z_\chi}$. The preliminary value of E158 is actually a little smaller than the Standard Model prediction by about 16\% (1.4 sigma). That leads to the 95\% CL constraint $m_{Z_\chi}\gsim1$ TeV\null.

What is the long term prospect for fixed target M\o ller scattering? An interesting possibility, suggested by K. Kumar at Snowmass (1996), is to use the 250--500 GeV polarized $e^-$ beam at a high energy $e^+e^-$ collider, but on a fixed target. Because of the higher energy and intensity, as well as the potential for much longer running, such an effort could reach an uncertainty $\Delta\sinsthw\mzms$ of about $\pm0.00006$. That would then represent the best determination of the weak mixing angle and a powerful constraint on New Physics. The only known way to do better is to measure $A_{LR}$ at the $Z$ pole with $\sim10^9 Z$ decays, using polarized $e^+$ and $e^-$ beams (one might attain $\Delta\sinsthw\mzms =\pm 0.00002$). However, a polarized $e^+$ beam is very technically challenging.

\section{Outlook and Conclusion}

The recent update of $m_t$ to $\sim178$ GeV renders the values of $m_W$ and $\sinslep$ very consistent within the Standard Model framework and together they imply a relatively light Higgs $\lsim 154$ GeV\null. That constraint indirectly suggests supersymmetry may be real and will soon be uncovered at the LHC\null. The 2.7 sigma discrepancy in $a^{\rm exp}_\mu - a^{\rm SM}_\mu$ can also be interpreted as a strong hint of supersymmetry.

It would be nice to clarify those hints by improved measurements of $\sigma(e^+e^-\to{\rm hadrons})$, $a^{\rm exp}_\mu$, $m_W$ and $\sinsthw\mzms$. The latter two may require a giga $Z$ facility and high statistics $e^+e^-$ running above the $W^+W^-$ threshold.

High precision low energy experiments such as atomic parity violation, neutrino scattering and polarized electron scattering also have a complementary role to play in constraining New Physics effects. However, it will be extremely difficult to push the current $\pm1\%$ uncertainty to $\pm0.1\%$, a challenging but appropriate long term goal.

Of course, high precision studies are only part of our future agenda. Thorough exploration of neutrino oscillations, including CP violation, search for edms and charged lepton flavor violation e.g.\ $\mu\to e\gamma$, $\mu^- N\to e^- N$, high energy collider probes and many other experiments will round out a progressive program of future discovery.

Unfortunately, one of the problems of our profession is time. It takes many years to propose, fund and complete new experiments (facilities). Most of us would like to see the process move much faster. However, progress eventually prevails. To illustrate the speedy ascent of particle physics, I list in table~\ref{tabtwo} the prevailing values of various fundamental parameters and progress in some areas as seen in 1993 \cite{marcssi93}, and 2004, years I lectured here at SLAC\null.

\begin{center}
\begin{table}[ht]
\caption{Changes from 1993--2004 \label{tabtwo}}
\begin{tabular}{lcc}
\\
Quantity & 1993 & 2004 \\[10pt]
$m_W$ (GeV) & $80.22\pm0.26$ & $80.426\pm0.034$ \\[6pt]
$\sinslep$ & 0.2318(6) & 0.23085(21) \\[6pt]
$m_t$ (GeV) & $>131$  & $178.0\pm4.3$ \\[6pt]
$m_H$ (GeV) & $57<m_H<800$ & $114<m_H<154$ \\[6pt]
$A_{LR}(e^-e^-)$ & An Impossible Idea & Experiment Completed \\[6pt]
$a_\mu$ & E821 Construction & Experiment Completed \\[6pt]
Neutrino Osc & A Speculation & Confirmed - Under Study \\[6pt]
Dark Energy & Einstein's Biggest Error & Believed 
\end{tabular}
\end{table}
\end{center} 

\noi Progress is slow but steady with some major advances along the way; but big questions remain. Why is top so heavy? What is $m_H$? Is the Higgs fundamental or composite? Why is parity (CP) violated? Does SUSY exist? What is dark matter? energy? Those types of provocative problems make particle physics stimulating and fun.

\bigskip

\leftline{Acknowledgement}
\medskip

\noi This work was supported by the High Energy Theory program
of the U.S. Department of Energy under Contract No. DE-AC02-98CH10886.

\end{document}